# Anomalous Hall and Nernst effects in ferrimagnetic $Mn_4N$ films: Possible interpretations and prospects for enhancement

Shinji Isogami[1*], Keisuke Masuda[1], Yoshio Miura[1], Rajamanickam Nagalingam[1], and Yuya Sakuraba[1,2]

[1] Research Center for Magnetic and Spintronic Materials, National Institute for Materials Science, Sengen 1-2-1, Tsukuba, Ibaraki 305-0047 Japan

[2] PRESTO, Japan Science and technology Agency, Saitama 332-0012, Japan

*E-mail: isogami.shinji@nims.go.jp

**Ferrimagnetic $Mn_4N$ is a promising material for heat flux sensors, based on the anomalous Nernst effect (ANE), because of its sizeable uniaxial magnetic anisotropy ($K_u$) and low saturation magnetization ($M_s$). We experimentally and theoretically investigated the ANE and anomalous Hall effect in sputter-deposited $Mn_4N$ films. It was revealed that the observed negative anomalous Hall conductivity ($\sigma_{xy}$) could be explained by two different coexisting magnetic structures, that is, a dominant magnetic structure with high $K_u$, contaminated by another structure with negligible $K_u$, owing to an imperfect degree of order of N. The observed transverse thermoelectric power ($S_{ANE}$) of +0.5 μV/K at 300 K yielded a transverse thermoelectric** **coefficient ($\alpha_{xy}$) of +0.34 A/(m·K), which was smaller than the value predicted from the first-principles calculation. The interpretation for $\alpha_{xy}$ based on the first-principles calculations led us to conclude that the realization of single magnetic structure with high $K_u$ and optimal adjustment of the Fermi level are promising approaches to enhance $S_{ANE}$ in $Mn_4N$ through the sign reversal of $\sigma_{xy}$ and the enlargement of $\alpha_{xy}$ up to a theoretical value of 1.77 A/(m·K).**

**<Main text>**

Thermoelectric conversion based on the anomalous Nernst effect (ANE)[1] is an intriguing alternative technology to those based on the conventional Seebeck effect,[2] because the transverse thermoelectric power based on the ANE ($S_{\mathrm{ANE}}$) is given in the perpendicular direction with respect to the heat current as shown in the following equation,

$$\boldsymbol{E}_{\mathrm{ANE}} = S_{\mathrm{ANE}} \boldsymbol{\nabla} T \times \left(\frac{\boldsymbol{M}}{|\boldsymbol{M}|}\right) , \qquad (1)$$

where $\boldsymbol{E}_{\mathrm{ANE}}$, $\boldsymbol{M}$, and $\boldsymbol{\nabla} T$ denote the electric field due to the ANE, magnetization, and temperature gradient, respectively. Owing to these characteristics, a prototype coplanar-type thermopile module has been fabricated using magnetic materials and an enhancement of the serial voltage has been demonstrated.[3,4]

One of the most promising applications of the ANE-based coplanar thermopile structure is heat flux sensors. Zhou *et al.* recently fabricated a prototype ANE-based heat flux sensor on a flexible thin polyimide sheet and successfully demonstrated heat flux sensing through ANE.[5]

Although the sensor has the potential to be versatile because of its high flexibility and low thermal resistance, its low sensitivity is the most important disadvantage that needs to be mitigated. To overcome this limitation, magnetic materials with large $S_{\mathrm{ANE}}$ are indispensable.[6] Furthermore, because the magnetization direction must be aligned to the width direction of magnetic wires, even at zero field, low $M_{\mathrm{s}}$ is also an important factor in decreasing the demagnetization field in magnetic wires that causes a reduction of $\boldsymbol{E}_{\mathrm{ANE}}$ in the zero-field state.[5] To stabilize the spontaneous magnetization, high uniaxial magnetic anisotropy ($K_{\mathrm{u}}$) is also desirable. In recent years, ferromagnetic Heusler alloys, such as $Co_2MnGa$[7-9] and $Co_2MnAl_{0.63}Si_{0.}37,10$) have been fabricated that exhibit high $S_{\mathrm{ANE}}$ of ~6 μV/K. However, their high $M_{\mathrm{s}}$ and low $K_{\mathrm{u}}$ are not suitable for a heat flux sensor. In contrast, 0.1~0.5 μV/K was reported in $Mn_3Sn$ films, irrespective of its extremely low $M_{\mathrm{s}}$ down to ~$10^{-3}$ T.[11] However, its $S_{\mathrm{ANE}}$ value is not sufficient for the sensor. Therefore, exploring magnetic materials with high $S_{\mathrm{ANE}}$, low $M_{\mathrm{s}}$, and high $K_{\mathrm{u}}$ has been of significant importance.

In the present study, we focus on $Mn_4N$ as a promising candidate material with low $M_{\mathrm{s}}$ and high $K_{\mathrm{u}}$. $Mn_4N$ film has long been known as a Mn-based perpendicular magnetic anisotropy (PMA) ferrimagnetic material with an antiperovskite structure described by the formula $ANB_3$, where A and B correspond to Mn(I) at corner sites and Mn(II) at face-centered sites, respectively.[12,13] Owing to the ferrimagnetism of $Mn_4N$, its $M_{\mathrm{s}}$ is much smaller than that in conventional ferromagnets, such as Co-based Heusler alloys, Fe-based alloys, and nitride systems. The high $K_{\mathrm{u}}$ in $Mn_4N$ is also an attractive feature for usage in ANE-based thermoelectric applications. The $Mn_4N$ films, fabricated by sputtering and molecular-beam epitaxy (MBE) in previous studies, exhibited $M_s \approx 0.1$ T and $K_u \approx 0.1$ MJ/m$^3$.[14-19] Although the AHE for $Mn_4N$ has been reported,[17,20] the ANE and AHE, including the analysis from the perspective of intrinsic mechanism, have not yet been studied.

In this study, we measured both the ANE and AHE of $Mn_4N$ films to analyze the mechanisms of ANE. Our results reveal that $Mn_4N$ is one of the few materials that satisfies the requirements of $M_s$ and $K_u$ for heat flux sensors, while its $S_{ANE}$ is moderate. The first-principles calculations revealed that two different magnetic structures coexist in present $Mn_4N$ films and adjustment of the Fermi level is necessary to enhance the ANE.

Two 20-nm-thick $Mn_4N$ films were grown on single-crystal MgO (001) substrates by reactive nitridation sputtering at substrate temperatures of 400 °C, 450 °C, and 500 °C (referred to as "sample-I," "sample-II," and "sample-III," respectively), and 2 nm thick Al films were then deposited as capping layers. The flow ratio of $N_2$ to Ar gas was 12 %. Structural analysis was performed using X-ray diffractometry (XRD) with Cu $K_\alpha$ radiation (SmartLab: Rigaku, Inc.). The magnetic properties were measured via a superconducting quantum interference device vibrating sample magnetometer (SQUID-VSM: Quantum Design, Inc.). The $K_u$ was estimated by $M_s H_k/2 + M_s^2/2\mu_0$, where the anisotropy field, $H_k$, is defined as the saturation magnetic field difference between the in-plane and out-of-plane directions. The ANE and AHE were measured by a physical property measurement system (VersaLab: Quantum Design, Inc.). Rectangular patterned samples with dimensions of $w$ = 2 mm and $l$ = 7 mm (where $w$ is width and $l$ is length) were prepared via photolithography and Ar-ion milling. The anomalous Hall (Nernst) voltage $V_{\mathrm{AHE}}$ ($V_{\mathrm{ANE}}$) was measured in the $y$-direction by applying $I_x$ ($\nabla T_x$) and the external magnetic field, $H_z$ ($H_z$), as shown in Fig. 1. $\nabla T_x$ was strictly measured by an infrared camera with a black body coating on the sample.[5,10,21] The Seebeck coefficients of the films ($S_{\mathrm{SE}}$) were measured by using a Seebeck coefficient or electric resistance measurement system (ZEM-3: ADVANCE RIKO, Inc.) equipped with R-type thermocouples.

In the theoretical calculation of the anomalous Hall conductivity ($\sigma_{xy}$) and the transverse thermoelectric coefficient ($\alpha_{xy}$), we combined the first-principles calculations and the Boltzmann transport theory. The electronic structure of the stoichiometric $Mn_4N$ was first calculated using the full-potential linearized augmented plane-wave method, including the effect of spin–orbit interactions, which was implemented in the WIEN2K program.[22] The lattice constants of the unit cell were fixed at $a$ = 3.87726 Å and $c$ = 3.83974 Å ($c/a$ = 0.99), which were the experimental values evaluated from the in-plane and out-of-plane XRD profiles. The self-consistent-field calculation using 24 × 24 × 25 $k$ points yielded both magnetic structures type-A and type-B. For the obtained electronic structures, we calculated $\sigma_{xy}$ and $\alpha_{xy}$ employing the same method as that used in our previous work.[9] In the calculation of $\alpha_{xy}$, the temperature $T$ was set to 300 K. We used 89 × 89 × 90 $k$ points for the Brillouin-zone integration, ensuring good convergence for $\sigma_{xy}$ and $\alpha_{xy}$.

Figure 2 shows the out-of-plane XRD profiles for sample-I, II, and III. As shown by dashed lines, the peak intensities were fitted with pseudo-Voigt functions.[23)] The diffraction peaks at $2\theta/\omega \approx 23°$ and 47° correspond to the (001) superlattice and (002) fundamental peaks of the $Mn_4N$ crystals, respectively. The degree of order (*S*) of fabricated $Mn_4N$ epitaxial films was estimated using the integral diffraction intensity of the $Mn_4N$ (001) and (002) peaks: $S = \sqrt{\frac{I_{001}^{\mathrm{obs}}/I_{002}^{\mathrm{obs}}}{I_{001}^{\mathrm{cal}}/I_{002}^{\mathrm{cal}}}}$, where $I_{hkl}^{\mathrm{obs}}$ and $I_{hkl}^{\mathrm{cal}}$ correspond to the integral intensities of the peaks, originating from (*hkl*) planes. These intensities were obtained from the pseudo-Voigt function fitting results and calculations, respectively (see Supplementary Material). The resultant *S* for sample-I, II, and III were estimated to be 0.6, 0.7, and 0.7, respectively, which were found to be relatively close to the film, deposited by MBE and sputtering techniques.[17,24)] Although an $Mn_4N$ unit cell, with a body-centered N, is the most stable phase, there are 12 possible equivalent sites for N atoms, similar to that in $Fe_4N$.[25)] Therefore, the *S* in this study represents the degree of order for both—N site ordering and its deficiency. Figure 2(b), 2(c), and 2(d) show the magnetization curves for sample-I, II, and III, respectively, where *H* was swept along the [100] and [001] directions of the $Mn_4N$ crystals. Comparing the two curves in different directions, the magnetic easy-axis was found to point in the [001] direction; that is, sample-II showed sizeable PMA ($K_u \approx 0.13$ MJ/m$^3$). The value of $\mu_0 M_s$ was measured to be 100 mT, comparable to that of $Mn_4N$ films fabricated by sputtering and MBE.[14-19)] In case of sample-I and III, the two magnetization curves, with $H$ || [100] and [001], showed similar magnetization process, indicating negligible PMA ($K_u \approx 0.003$

and 0.001 MJ/m$^3$ for sample-I and -III, respectively). The value of $\mu_0 M_s$ was decreased to 90 and 60 mT for sample-I and -III, respectively. These crystal structure and magnetization results show that 450 °C is an optimum substrate temperature for $Mn_4N$ to yield high PMA with higher $S$, in this study.

From the linear response equation, $S_{ANE}$ is given by[9)]

$$S_{\mathrm{ANE}} = \rho_{xx}\alpha_{xy} + \rho_{xy}\alpha_{xx} \ , \qquad (2)$$

where $\rho_{xx(xy)}$ and $\alpha_{xx}$ denote the longitudinal (anomalous Hall) resistivity and longitudinal thermoelectric coefficients, respectively. The second term of Eq. (2) corresponds to the contribution of the AHE to the ANE; therefore, the measurement of $\rho_{xy}$ is necessary to study the origin of $S_{ANE}$. Figure 3(a) shows the $\rho_{xy}$ curves for sample-I, II, and III as functions of $H_z$ at 300 K, indicating hysteresis similar to the $M$–$H$ curves shown in Fig. 2(b), 2(c), and 2(d). Figure 3(b) and 3(c) show the measurement temperature ($T$) dependence of $\rho_{xy}$ and $\rho_{xx}$, respectively. $\rho_{xx}$ decreased with decreasing $T$, suggesting metallic properties in the present three films. Figure 3(d) shows the $T$ dependence of $\sigma_{xy}$, estimated using the following formula:[9)]

$$\sigma_{xy} = -\frac{\rho_{xy}}{\rho_{xy}^2+\rho_{xx}^2} \ . \qquad (3)$$

$\sigma_{xy}$ was negative for both films: $\sigma_{xy} \approx -50$, $-80$, and $-78$ S/cm at 300 K, and $\sigma_{xy} \approx -10$, $-100$, and $-98$ S/cm at 4 K for sample-I, II, and III, respectively. The inset of Fig. 3(d) represents the $\sigma_{xy}$–$\sigma_{xx}$ relationship for the three samples. $\sigma_{xy}$ for sample-II and III remains constant in contrast to that of sample-I, which qualitatively suggests that sample-II and III contain stronger intrinsic contribution of AHE, compared with sample-I.

Figure 4(a) shows the $H_z$ dependence of $V_{ANE}$ normalized by $w$ and $\nabla T_x$ for the three samples. To evaluate $S_{ANE}$, $V_{ANE}$ was measured under different $\nabla T_x$ and the intrinsic component of ANE was evaluated by linear curve fitting in the high $\mu_0 H_z$ region and taking the intercept at zero field. The inset shows the plot of the dependence of $V_{ANE}/w$, obtained by the linear fitting on $\nabla T_x$ for sample-II; the slope of least-squares fitting yields an averaged $S_{ANE}$ of 0.50 μV/K. The $\alpha_{xy}$ for sample-II was estimated to be 0.34 A/(m·K), using Eq. (2), with the measured values of $S_{ANE}$, $\rho_{xx(xy)}$, and longitudinal thermoelectric coefficient, $\alpha_{xx}$ (= $S_{SE}/\rho_{xx}$). In contrast, the $\alpha_{xy}$ for sample-I and II were estimated to be 0.16 and 0.21 A/(m·K), respectively. Figure 4(b) shows the reported $S_{ANE}$ for various materials plotted against $M_s$. Based on the analysis of the required $\mu_0 M_s$ and $S_{ANE}$ for an ANE-based heat flux sensor, reported by Zhou *et al.*,[5] $\mu_0 M_s < 0.2$ and $S_{ANE} > 10$ μV/K are required to suppress the demagnetization field to less than 0.1 T and the sensitivity to 10 μV/(Wm$^{-2}$) in a 1 × 1 cm$^2$ size sensor with 10 μm as both height and the width of the magnetic wires. Therefore, this required range is highlighted in Fig. 4(b). Although the observed $S_{ANE}$ for sample-II [marked by a star in Fig. 4(b)] does not reach the target region showing the necessary values for heat flux sensors, one can notice that $Mn_4N$ is one of the few materials with small $M_s$ that satisfies this target and shows a relatively high $S_{ANE}$ value, as compared to other materials, such as $Mn_3Sn$[11] and $DO_{22}$-$Mn_3Ga$.[26]

The first-principles calculations for $\sigma_{xy}$ and $\alpha_{xy}$ were carried out from the perspective of intrinsic mechanism to understand the origin of observed AHE and ANE in the $Mn_4N$ films. Two possible collinear magnetic structures for $Mn_4N$ were investigated in our previous study (see Fig. 1 for the magnetic structures): "type-A," in which the magnetic moments of Mn(I) and Mn(II) point in opposite directions, giving tiny $M_s$ (≈ 7.9 mT) and negligible $K_u$; and "type-B," in which the magnetic moments couple parallel to each other, within the *ab*-plane, and alternately, along the *c*-direction, giving low $M_s$ (≈ 0.18 T) and high $K_u$ (≈ 4 MJ/m$^3$).[19] Type-B is energetically more stable, compared with type-A, when $S = 1$.[19] Because $\mu_0 M_s$ was measured to be 100 mT and a clear PMA ($K_u$ ≈ 0.13 MJ/m$^3$) was observed, we inferred that type-B magnetic structure was dominant in sample-II. Therefore, first, the calculation of $\sigma_{xy}$ with type-B was performed [solid line in Fig. 5(a)]. It was revealed that the calculated $\sigma_{xy}$ of 573 S/cm at the Fermi level ($\mu = 0$ eV) was inconsistent with the experimental value of –100 S/cm at 4 K. Such a large discrepancy motivated us to calculate $\sigma_{xy}$ with type-A structure [dashed line in Fig. 5(a)]. As a result, we found a negative $\sigma_{xy}$ at the Fermi level ($\sigma_{xy}$ = –500 S/cm), which agreed with the sign of the experimental value, although its magnitude still showed a discrepancy. However, notably, our previous study showed tiny $M_s$ and negligible $K_u$ for the type-A structure,[19] which indicates that type-A magnetic structure cannot be dominant in sample-II. These results imply that the main magnetic structure for sample-II is type-B; however, it is contaminated by a certain amount of type-A, whose contribution to AHE is large enough to reverse the sign of $\sigma_{xy}$ to negative in sample-II. Although the cause of type-A contamination is not fully understood, imperfect $S$ should be considered as a probable reason for such contamination. Our previous investigation revealed that $S = 1$ can contribute towards forming type-B rather than type-A due to the energy stability of the magnetic structure.[19)] Thus, we can infer that type-A—a second stable magnetic structure—possibly appears when $S$ deviates from 1. In contrast to our samples, Kabara *et al.* measured positive $\sigma_{xy}$ at 4 K for 100 nm thick $Mn_4N$ films, the sign of which agreed with that for type-B.[27)] Such a consistency can be attributed to the enhanced $S$ ($\approx 0.85$) due to the large layer thickness,[17)] and the contribution from type-A, if any, may be vailed by that from type-B. As for the experimental results shown in Fig. 2(b) and 2(c), the $M_s$ and $K_u$ for sample-I ($S = 0.6$) were less than those for sample-II ($S = 0.7$), which could be related to the higher contamination of type-A due to lower $S$ in sample-I. However, the $M_s$ and $K_u$ for sample-III were smaller than those for sample-II, as shown in Fig. 2(c) and 2(d), in spite of the similar, $S$, i.e., $S = 0.7$. Although these magnetic properties can be also attributed to the contamination of type-A, additional mechanisms are considered because the coercivity for sample-III clearly increased, as compared to sample-II. We observed island-like surface morphology and/or large surface roughness in sample-III (See supplementary material), via atomic force microscopy (AFM). These structural characteristics are responsible for the magnetic properties observed in sample-III through pinning of domain walls, as reported in the case of Co/Ni multilayer.[28)] On the other hand, the negatively smaller $\sigma_{xy}$ for sample-I, compared with those for sample-II and III, cannot be explained by the higher type-A contamination, because type-A contribution decreases the entire $\sigma_{xy}$. We can see a steep change in the calculated $\sigma_{xy}$ around the Fermi level in both type-A and –B structures, suggesting that $\sigma_{xy}$ is sensitive to $\mu$ [Fig. 5(a)]. For example, the change in $\mu$, within ±0.05 eV, results in a large variation width, of ~400 S/cm, for both magnetic structures. Therefore, the $S$-dependent shift in the Fermi level can be another cause of the negatively smaller $\sigma_{xy}$ for sample-I, as compared to those of sample-I and III.

Figure 5(b) shows the $\mu$ dependence of $\alpha_{xy}$, where $T$ is set to 300 K. At the Fermi level, $\mu = 0$, $\alpha_{xy} = 1.66$ (1.77) A/(m·K) for type-A (type-B). Comparing $|\alpha_{xy}|$ for the representative materials, we obtain $|\alpha_{xy}| = 4$ for $Co_2MnGa$,[7)] 1.2 for FeGa,[21)] 1.1 for MnSi,[29)] 0.6 for $SrRuO_3$,[30)] and 0.3 for $Mn_3Sn$[11)] from the experiments, and 0.9 for FePt,[31)] 0.52 for Co,[31)] and 0.21 for Fe[31)] from the calculations [all in units of A/(m·K)]. From these values, one can find that the calculated $\alpha_{xy}$ for $Mn_4N$, based on intrinsic origins, is sufficiently high. Therefore, $Mn_4N$ can be a promising material for ANE. Unlike $\sigma_{xy}$, high $\alpha_{xy}$ with a positive sign was predicted for both magnetic structures, which is beneficial for the enhancement of $S_{ANE}$, even in the case of type-A contamination. However, the measured $\alpha_{xy}$ was one order of magnitude smaller than that from the calculation. One possible explanation is the shift of Fermi level from the theoretically obtained value as well as its effect on $\sigma_{xy}$, because the peak around the Fermi level is remarkably sharp [Fig. 5(b)]. For example, the even small change in $\mu$ with ±0.05 eV provides a large variation width of ~2.7 (~1.8) A/(m·K) for type-A (type-B). The $S$ in $Mn_4N$ is related to the site ordering of N as well as the N deficiency (shown in Fig. 2); therefore, less $S$ corresponds to a decrease in the number of electrons in the $Mn_4N$ unit cell. This can be responsible for the shift in the Fermi level, on the basis of the rigid-band model.[32] In addition, quantifying the extrinsic contributions, such as skew-scattering and/or side-jump, may be necessary for explaining the dicrepancy,[33] although the nearly constant $\sigma_{xx} \approx 5 \times 10^3$ $(\Omega \text{ cm})^{-1}$ for sample-II and III are categorized into the intrinsic regime, based on the $\sigma_{xy}$–$\sigma_{xx}$ relationship,[34] as shown in the inset of Fig. 3(d).

Next, we address how to boost the ANE in $Mn_4N$. Equation (3) states that $S_{\mathrm{ANE}}$ decreases and increases when the signs of the first ($\rho_{xx}\alpha_{xy}$) and second ($\rho_{xy}\alpha_{xx}$) terms are opposite and same, respectively. In the case of the present $Mn_4N$ films, as shown in Table I, the resultant $S_{\mathrm{ANE}}$ lost due to the negative sign of $\rho_{xy}\alpha_{xx}$. In addition, the enhanced $\alpha_{xy}$, which is realized via modulation of the Fermi level, is also necessary to boost $S_{\mathrm{ANE}}$ for a $Mn_4N$ system. To modulate the Fermi level and obtain positive $\rho_{xy}\alpha_{xx}$, realizing $Mn_4N$ with ideal type-B and no or negligible contamination from type-A, via improved $S$, is a promising approach. Furthermore, the substitution of a third element, such as Ni, into $Mn_4N$ may be another interesting approach for enhancing $S_{\mathrm{ANE}}$ as well as reducing $M_{\mathrm{s}}$. This is because the sign reversal of AHE occurs with a Ni content of ~5 at.%.[35] Such a small Ni content retains the antiperovskite crystal structure,[35] suggesting a similar band structure and modulation of the Fermi level, based on the rigid-band model.[32] Therefore, the results of the present study can open a pathway for fabricating $Mn_4N$-based materials to develop heat flux sensors.

**Table I.** Summary of coefficients related to the ANE and AHE for $Mn_4N$.

| | Experiment | | | Calculation | |
|---|---|---|---|---|---|
| | Sample-I (400 °C) | Sample-II (450 °C) | Sample-III (500 °C) | Type-A | Type-B |
| $\mu_0 M_s$ (mT) | 88 | 100 | 60 | 7.9 (0 K) [19)] | 180 (0 K) [19)] |
| $K_u$ (MJ/m$^3$) | 0.003 | 0.13 | 0.001 | ~0 (0 K) | 4 (0 K) |
| $\rho_{xx}$ (μΩ cm) | 240 | 190 | 230 | | |
| $\rho_{xy}$ (μΩ cm) | 2.6 | 2.8 | 3.8 | | |
| $\sigma_{xy}$ (S/cm) | –50<br>–10 (4 K) | –80<br>–100 (4 K) | –78<br>–98 (4 K) | –573 (0 K) | 500 (0 K) |
| $\theta_{AHE}$ (%) | 1.1 | 1.5 | 1.65 | | |
| $S_{SE}$ (μV/K) | –14 | –9.7 | –11.8 | | |
| $S_{AHE} = S_{SE} \tan\theta_{AHE}$ (μV/K) | –0.15 | –0.15 | –0.19 | | |
| $\alpha_{xx}$ [A/(m·K)] | –5.7 | –5.1 | –5.0 | | |
| $\alpha_{xy}$ [A/(m·K)] | 0.16 | 0.34 | 0.21 | 1.77 | 1.66 |
| $\rho_{xx}\alpha_{xy}$ (μV/K) | 0.38 | 0.65 | 0.49 | | |
| $\rho_{xy}\alpha_{xx}$ (μV/K) | –0.15 | –0.15 | –0.19 | | |
| $S_{ANE}$ (μV/K) | 0.23 | 0.50 | 0.30 | | |

In summary, we studied the AHE and ANE for sputter-deposited 20-nm-thick ferrimagnetic $Mn_4N$ films. $S_{ANE}$ was measured to be 0.50 μV/K, which is the maximum value at 300 K, demonstrating that $Mn_4N$ is one of the few materials that shows relatively large $S_{ANE}$ with low $M_s$ and high $K_u$. The $Mn_4N$ film characteristics, revealed via the first-principles calculations in this study, are: i) the possible magnetic structure of the film may be type-B; however, type-A, with small $M_s$ and negligible $K_u$, may be partly included; ii) $S_{ANE}$ decreases because of the imperfect $S$, thereby reducing $\alpha_{xy}$ and causing the negative sign of the AHE component. To reach the target of heat flux sensors, further enhancement of $S_{ANE}$ of $Mn_4N$ is critical. Useful approaches to be considered are: obtaining a positive $\sigma_{xy}$ and enhanced $\alpha_{xy}$ via the Fermi-level tuning and improving $S$, leading to type-B magnetic structure. Thus, the present study provides insights into the fundamentals of the transverse transport properties of Mn-based nitride systems.

**<Supplementary Material>**

See supplementary material for details on estimating the degree of order of $Mn_4N$ and the surface morphologies, observed via AFM.

**<Acknowledgement>**

This work was supported by JSPS KANENHI via grant no. 19K04499, 19K21954, 18H03787, and 16H06332; PRESTO from the Japan Science and Technology Agency (No. JPMJPR17R5); and NEDO "Mitou challenge 2050" (Grant No. P14004). The authors thank B. Masaoka for his technical support.

Part of this work was carried out under the Cooperative Research Project Program of the RIEC, Tohoku University.

## <Data Availability>

The data that support the findings of this study are available from the corresponding author upon reasonable request.

**<References>**

**<Figure captions>**

**Figure 1.** Measurement configuration of anomalous Nernst voltage ($V_{ANE}$) and anomalous Hall voltage ($V_{AHE}$) for the MgO substrate/$Mn_4N$ (20 nm)/Al (2 nm). The right schematic illustration represents the crystal structure of the $Mn_4N$ unit cell and the magnetic structures, type-A and type-B.

**Figure 2.** (a) Out-of-plane X-ray diffraction profiles of sample-I (substrate temperature, $T_{sub}$ = 400 °C), sample-II (450 °C), and sample-III (500 °C). The dashed lines indicate the peak fitting to evaluate the degree of order of N ($S$). (c, d, e) Magnetization curves of sample-I, sample-II, and sample-III.

**Figure 3.** (a) Representative anomalous Hall resistivity ($\rho_{xy}$) as a function of the applied magnetic field along $z$-direction ($H_z$) at 300 K. (b, c, d) Dependence of anomalous Hall resistivity ($\rho_{xy}$) (b), longitudinal resistivity ($\rho_{xx}$) (c), and anomalous Hall conductivity ($\sigma_{xy}$) (d) on measurement temperature ($T$). The inset represents the plot of $|\sigma_{xy}|$ vs $\sigma_{xx}$.

**Figure 4.** (a) Representative anomalous Nernst voltage ($V_{ANE}$) at 300 K as a function of the applied magnetic field, pointing in the $z$-direction ($H_z$). The inset represents saturated $V_{ANE}$ as a function of temperature gradient ($\nabla T_x$). The solid line shows the least-squares fitting with a linear function. The corresponding slope yields $V_{ANE}/(w\cdot \nabla T_x)$. (b) Thermoelectric power ($S_{ANE}$), for various materials,

plotted against saturation magnetization ($\mu_0 M_s$). Red color represents the target values for heat flux sensors.

**Figure 5.** (a,b) Calculated anomalous Hall conductivity ($\sigma_{xy}$) (a) and transverse thermoelectric coefficient ($\alpha_{xy}$) (b) as functions of chemical potential ($\mu$) for the $Mn_4N$ unit cell. The dashed and solid lines represent the results for magnetic structure type-A and type-B, respectively.

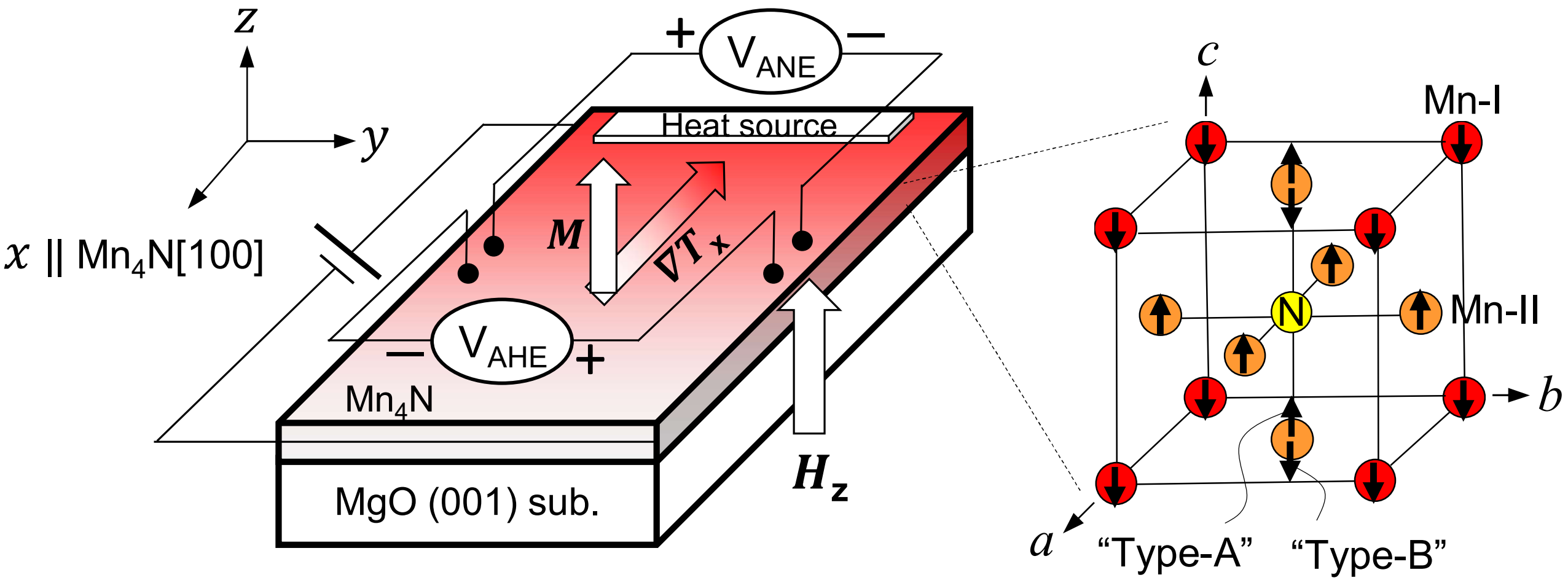

Fig. 1

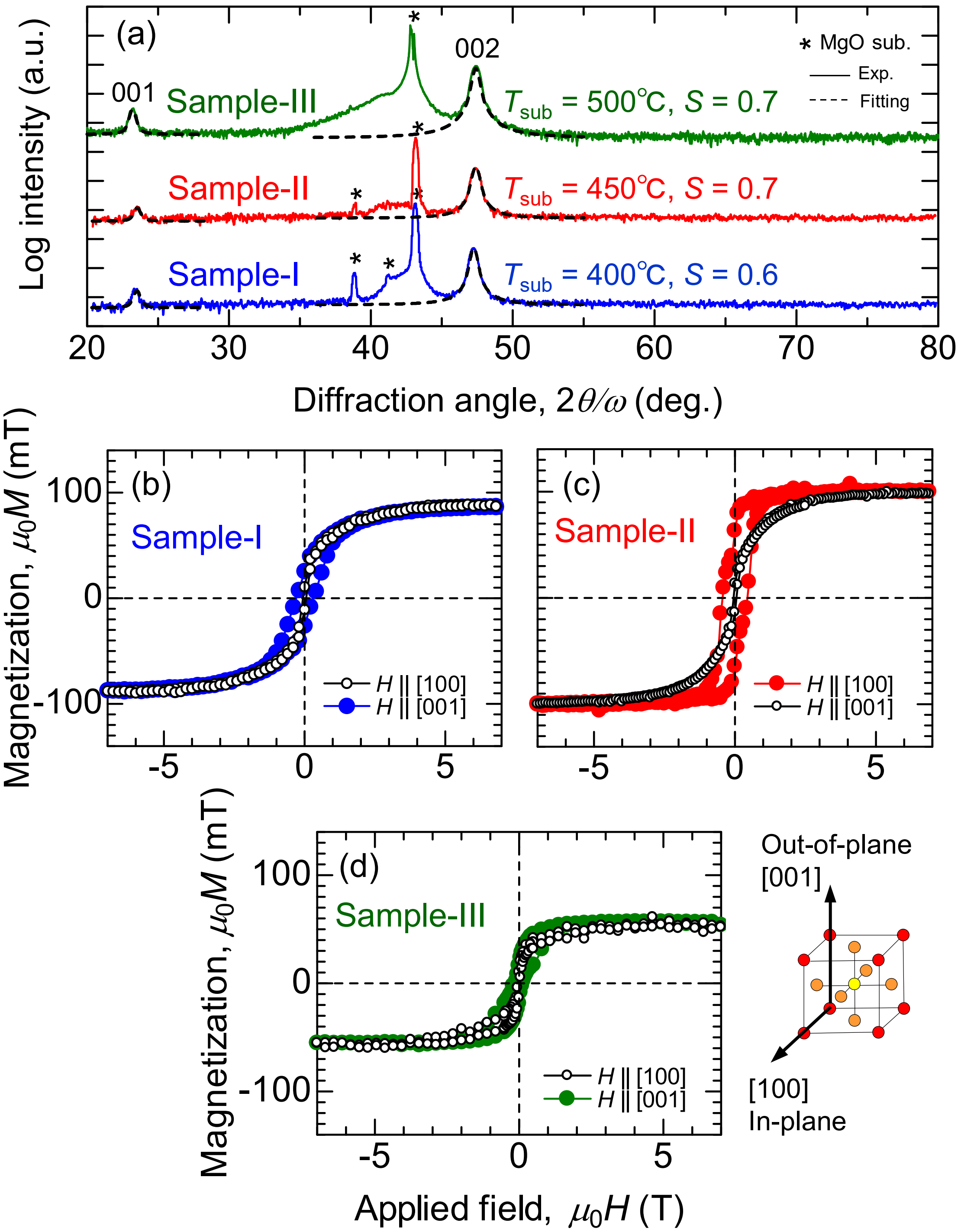

Fig. 2

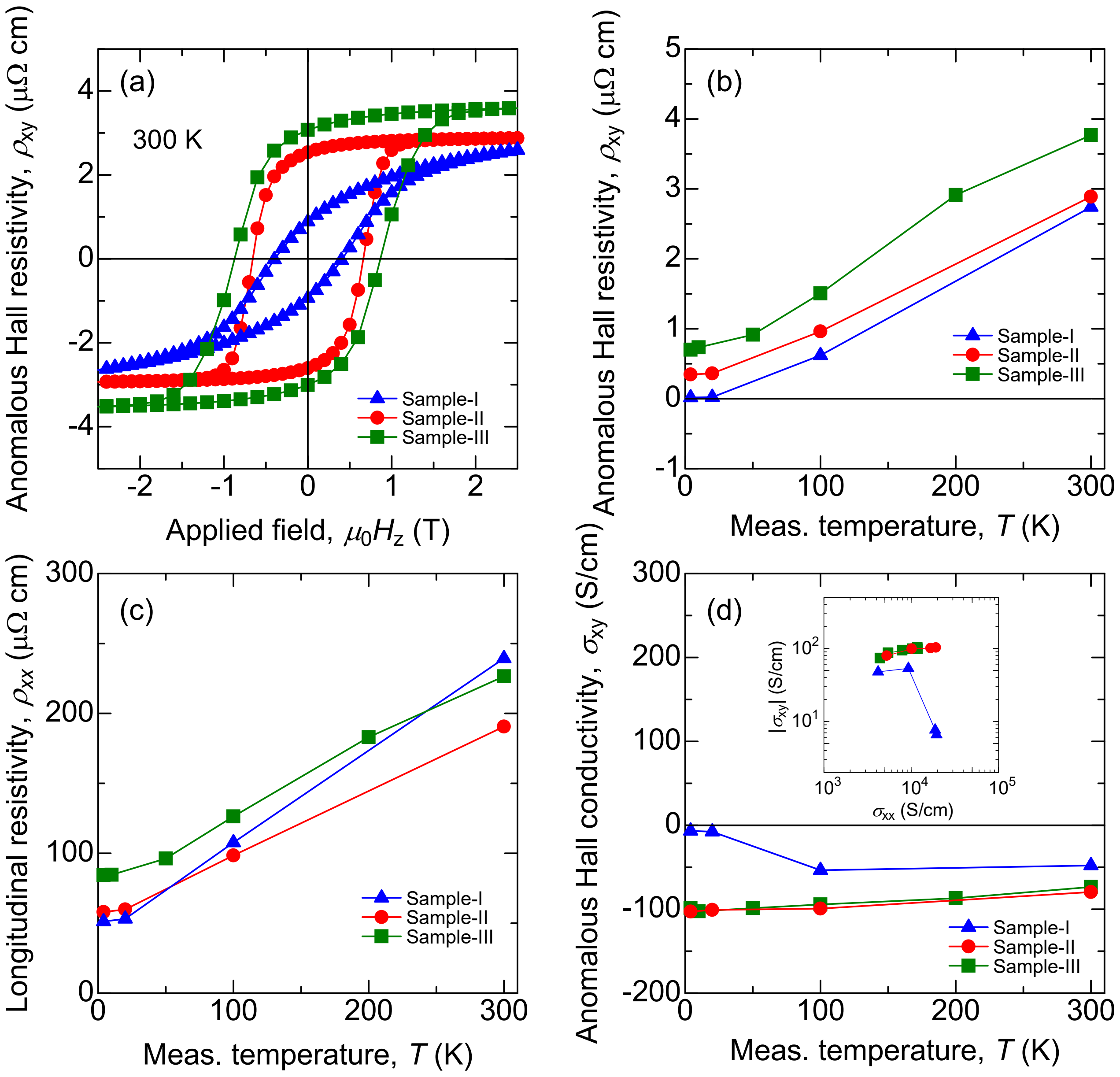
(a)
300 K
Anomalous Hall resistivity, ρxy (μΩ cm)
Applied field, μ0Hz (T)
Sample-I
Sample-II
Sample-III
(b)
Anomalous Hall resistivity, ρxy (μΩ cm)
Meas. temperature, T (K)
(c)
Longitudinal resistivity, ρxx (μΩ cm)
Meas. temperature, T (K)
(d)
Anomalous Hall conductivity, σxy (S/cm)
|σxy| (S/cm)
σxx (S/cm)
Meas. temperature, T (K)

Fig. 3

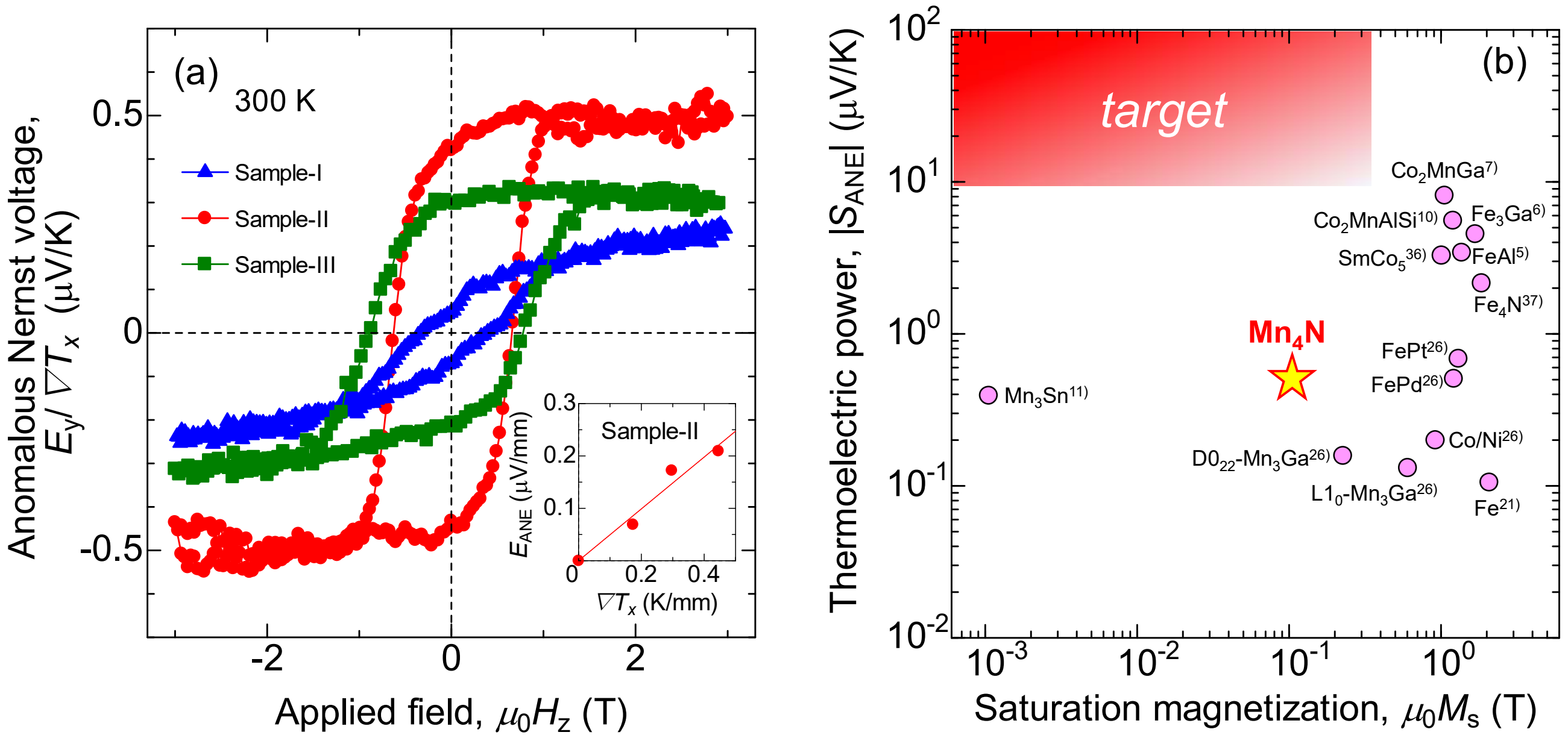

Fig. 4

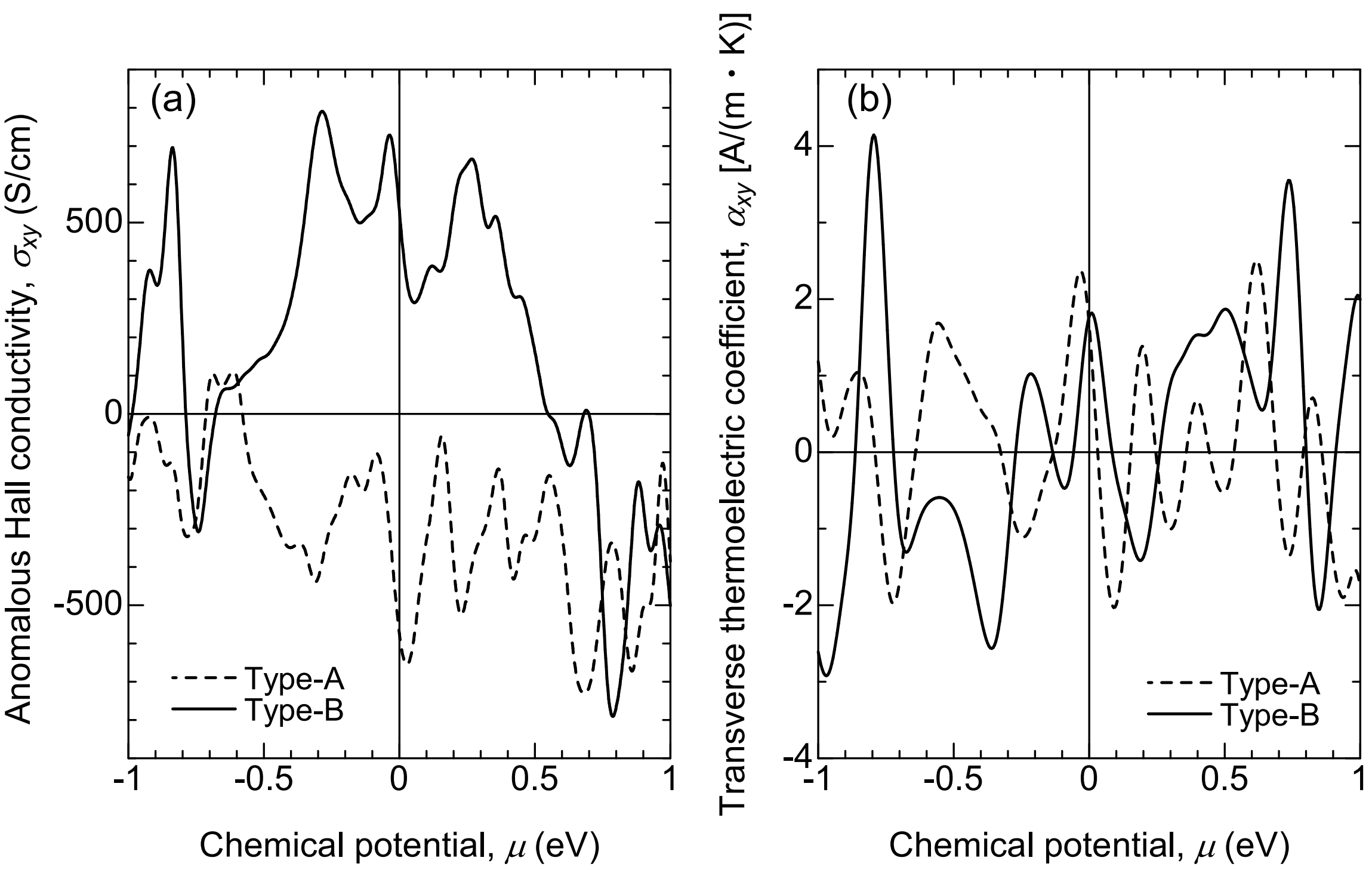
(a)
Anomalous Hall conductivity, $\sigma_{xy}$ (S/cm)
500
0
-500
Type-A
Type-B
-1
-0.5
0
0.5
1
Chemical potential, $\mu$ (eV)
(b)
Transverse thermoelectric coefficient, $\alpha_{xy}$ [A/(m · K)]
4
2
0
-2
-4
Type-A
Type-B
-1
-0.5
0
0.5
1
Chemical potential, $\mu$ (eV)

Fig. 5

Supplementary Material

# Anomalous Hall and Nernst effects in ferrimagnetic $Mn_4N$ films: Possible interpretations and prospects for enhancement

Shinji Isogami[1a)], Keisuke Masuda[1], Yoshio Miura[1], Rajamanickam Nagalingam[1], and Yuya Sakuraba[1,2]

[1] National Institute for Materials Science, Tsukuba 305-0047, Japan

[2] PRESTO, Japan Science and technology Agency, Saitama 332-0012, Japan

a) isogami.shinji@nims.go.jp

## S1. Estimation of degree of order (*S*)

From the XRD profiles, the degree of order (*S*) of the fabricated $Mn_4N$ epitaxial films was estimated as:

$$S = \sqrt{\frac{I_{001}^{\mathrm{obs}}/I_{002}^{\mathrm{obs}}}{I_{001}^{\mathrm{cal}}/I_{002}^{\mathrm{cal}}}} \ , \quad \text{(S1)}$$

$$I_{hkl}^{\mathrm{cal}} = A \cdot |F_{hkl}|^2 \cdot LP \ , \quad \text{(S2)}$$

where

$$A = \frac{1}{2\mu}\left(1 - e^{-2\mu t/\sin\theta}\right) \ , \quad \text{(S3)}$$

$$LP = \frac{1+\cos^2(2\theta)\cos^2(2\theta_M)}{\sin(2\theta)\{1+(\cos^2\theta_M)\}} \ . \quad \text{(S4)}$$

$I_{hkl}^{\mathrm{obs}}$ and $I_{hkl}^{\mathrm{cal}}$ correspond to the integral peak intensities of peaks from (*hkl*) planes, which are obtained by the pseudo-Voigt function fitting results and calculations, respectively. *A* and *LP* correspond to the absorption factor with line absorption coefficient ($\mu$) and film thickness ($t$), the Lorenz polarization factor with Bragg angle ($\theta_M$) of the monochromator Ge(220) = 22.65°, respectively. The structure factor ($F_{hkl}$) of $Mn_4N$ was estimated using:

$$F_{hkl} = f_{\mathrm{Mn}}\{1 + (-1)^{h+k} + (-1)^{k+l} + (-1)^{l+h}\} + f_{\mathrm{N}}(-1)^{h+k+l} \ , \quad \text{(S5)}$$

where $f$ represents the atomic scattering factor.

## S2. Surface morphology of the 20-nm-thick $Mn_4N$ film with higher substrate temperature ($T_{sub}$ = 500 °C)

Figure S1 shows a plot of the surface roughness ($R_a$) as a function of the substrate temperature ($T_{sub}$) for the stacked sample: MgO substrate//$Mn_4N$ (20 nm)/Al (2 nm). $R_a$ shows similar values at $T_{sub}$ = 400 °C and 450 °C; however, it increases by 50 % at $T_{sub}$ = 500 °C. The AFM images show that island-like grains appear at $T_{sub}$ = 500 °C, which may be responsible for the increased $R_a$.

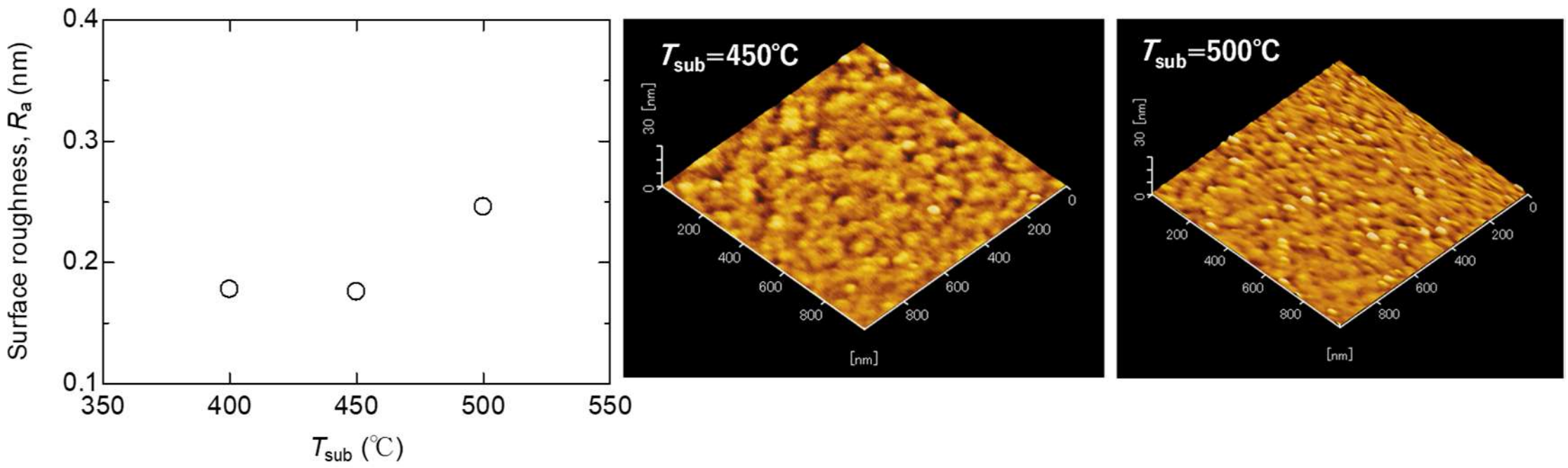

Figure S1 Surface roughness ($R_a$) as a function of the substrate temperature ($T_{sub}$) for the stacking—MgO substrate//$Mn_4N$ (20 nm)/Al (2 nm). Both the surface morphologies were observed via AFM with 1 μm × 1 μm in dimension.